\documentclass[%
 aps,prl,
 amsmath,amssymb,
reprint,%
]{revtex4-1}

\usepackage{graphicx}% Include figure files
\usepackage{dcolumn}% Align table columns on decimal point
\usepackage{bm}% bold math
\usepackage{rotating}
\usepackage[english]{babel}

\newcommand{\rr}{\mathbf{r}}

\newcolumntype{m}{D{.}{.}{1.4}}
\newcolumntype{n}{D{.}{.}{2.4}}
\renewcommand{\H}{\ensuremath{\text{H}}}

\newcommand{\lr}{\ensuremath{\text{lr}}}
\newcommand{\sr}{\ensuremath{\text{sr}}}

\begin{document}

\title{
Combining Density Functional Theory and Density Matrix Functional Theory}

\author{Daniel R. Rohr$^1$}
\author{Julien Toulouse$^{2}$}
\author{Katarzyna Pernal$^{1}$}
\affiliation{$^{1}$Institute of Physics,
Technical University of \L \'od\'z,
W\'olcza\'nska 219,
93-005 \L\'od\'z, Poland}%
\affiliation{$^{2}$Laboratoire de Chimie Th\'eorique,
Universit\'e Pierre et Marie Curie and CNRS,
4 place Jussieu,
75252 Paris, France}

\date{\today}% It is always \today, today,

\begin{abstract}
We combine density-functional theory with density-matrix functional theory to get the best of both worlds. This is achieved by range separation of the electronic interaction which permits to rigorously combine a short-range density functional with a long-range density-matrix functional. The short-range density functional is approximated by the short-range version of the Perdew-Burke-Ernzerhof functional (srPBE). The long-range density-matrix functional is approximated by the long-range version of the Buijse-Baerends functional (lrBB). The obtained srPBE+lrBB method accurately describes both static and dynamic electron correlation at a computational cost similar to that of standard density-functional approximations. This is shown for the dissociation curves of the H$_{2}$, LiH, BH and HF molecules.
\end{abstract}
%Use showkeys class option if keyword
\maketitle

Density-functional theory (DFT)~\cite{Hohenberg:1964p65,KohSha-PR-65} is a widely-used approach for electronic-structure calculations in quantum chemistry and condensed-matter physics. In particular, for molecular systems, its success lies in the fact that the common density-functional approximations (e.g., semilocal and hybrid functionals) give reasonably accurate thermodynamical properties near equilibrium geometries, at low computational cost. Indeed, most DFT implementations have a computational cost that scales at worst as $M^4$, where $M$ is the number of basis functions, and yield results with a weak dependence of the basis size. 

The accuracy of common density-functional approximations near equilibrium geometries is usually attributed to the correct description of short-range dynamic electron correlation (see, e.g., Ref.~\onlinecite{GriSchBae-JCP-97}). Another form of electron correlation is the so-called static (or strong) correlation, which is present in systems with electrons occupying partially-filled nearly-degenerate states. Examples of systems with static correlation are transition metals and systems with partially or fully broken bonds. The usual density-functional approximations most often fail to adequately describe this type of correlation (see, e.g., Refs.~\onlinecite{Sav-INC-96,Bae-PRL-01,CohMorYan-JCP-08}). This is unfortunate since for example bond cleavage is an ubiquitous process for chemistry.

Density-matrix functional theory (DMFT) (see, e.g., Refs.~\onlinecite{GILBERT:1975p148,Muller:1984p1,GoeUmr-PRL-98,Buijse:2002p30,Gritsenko:2005p108,Rohr:2008p59}) has recently emerged as a promising alternative to overcome the limitations of usual DFT approaches. The energy is expressed as a functional of the one-electron reduced density matrix. The use of the density matrix provides more flexibility beyond single-determinant DFT. It offers an explicit description of static correlation by fractional occupation numbers for the orbitals. A few density-matrix functional approximations were successful in describing bond dissociation curves of small test systems~\cite{Gritsenko:2005p108,Rohr:2008p59}. However, these functionals are computationally more demanding than usual density-functional approximations. They rely on a transformation of the two-electron integrals, which makes them scale with $M^{5}$. Moreover, DMFT generally suffers from a strong basis-size dependence.

One of the earliest DMFT approximation is the Buijse-Baerends (BB) functional (also called Corrected-Hartree, or M\"uller functional)~\cite{Buijse:2002p30,Muller:1984p1}. Its computational cost compares to that of usual density functional approximations, since it also scales with $M^{4}$ and an efficient optimization scheme is available~\cite{Cances:2008p2948}. The spin-restricted BB functional captures the essence of static correlation in bond dissociation, as indicated by the physically correct saturation of its total energy at reasonable bond distances. In contrast, spin-restricted density functional approximations yield total energies that keep increasing at unreasonably large distances. However, the BB total energy is much too low, which suggests that it poorly describes dynamic correlation.

In this Letter, we present the first molecular tests of a theory which combines DFT and DMFT to get the best of both worlds. The method is based on the range-separation scheme (see Ref.~\onlinecite{Stoll:1985p177} for the original idea, and e.g. Ref.~\onlinecite{Toulouse:2004p062505} for details) which permits rigorous combination of a short-range density functional with a long-range density-matrix functional~\cite{Pernal:2010p052511}. The idea of this theory, which we name srDFT+lrDMFT, is that dynamic correlation should be mostly described by the short-range density functional, while static correlation should be mostly accounted for by the long-range density-matrix functional. In principle, the theory is exact, as standard DFT and DMFT. In practice, we use the short-range version of the Perdew-Burke-Ernzerhof density functional (srPBE) of Ref.~\onlinecite{Goll:2005p2378}, and the long-range version of the BB density-matrix functional (lrBB)~\cite{Pernal:2010p052511}, which has already been successfully tested for the homogeneous electron gas. The obtained srPBE+lrBB method describes accurately both dynamic and static correlation, as demonstrated for the dissociation curves of several test systems. Moreover, the method has a computational cost, which compares to common DFT methods. It scales as $M^{4}$ and has a weak basis-size dependence.

\begin{figure}
\includegraphics[scale=0.35]{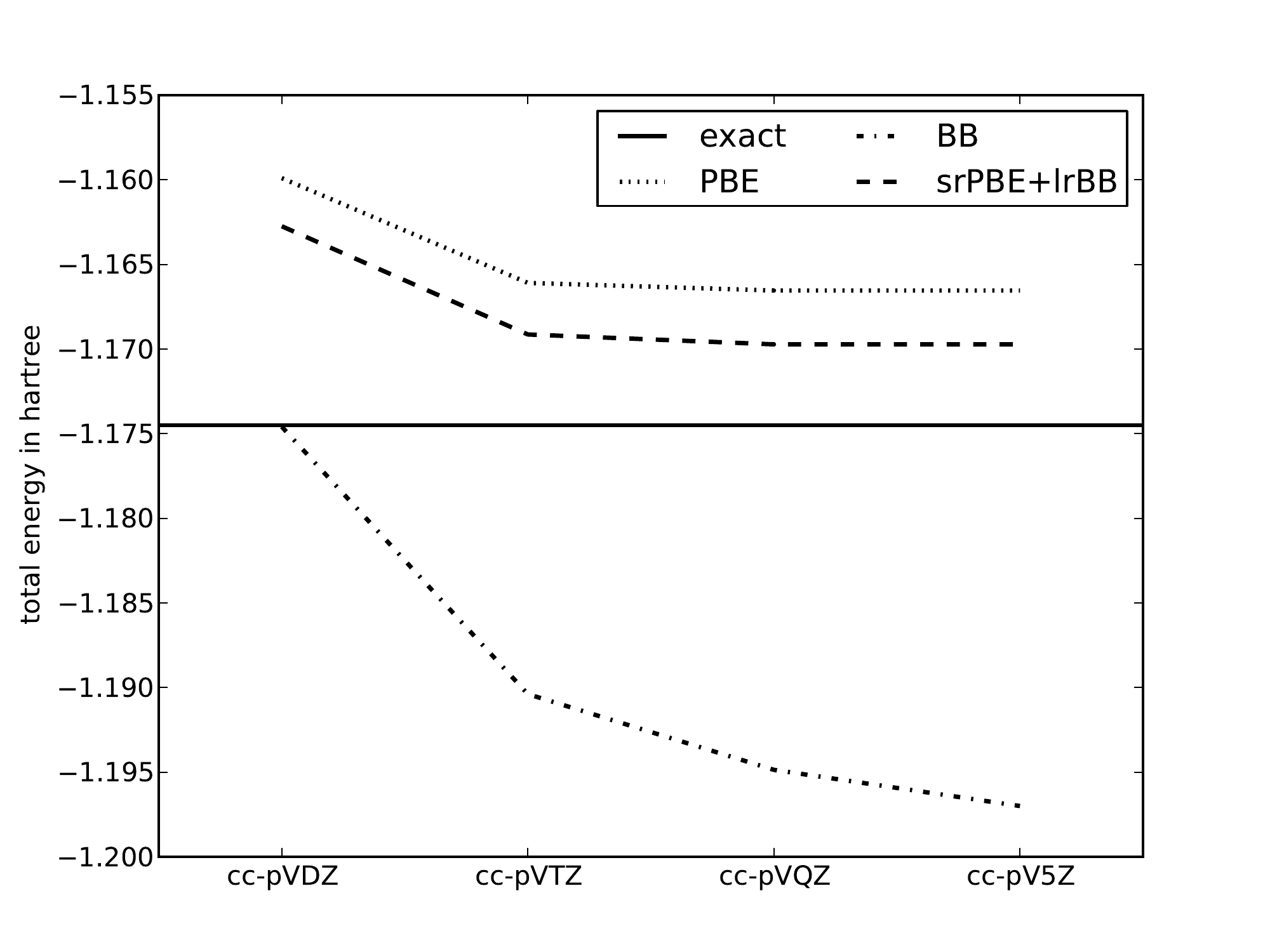}
\caption{\label{fig:bs}Basis set convergence for the H$_{2}$ molecule at the equilibrium distance (1.401 bohr).}
\end{figure}

\vskip 2mm \noindent {\it Theory.} \hskip 2mm
We now derive the equations of the srDFT+lrDMFT theory.
Following Hohenberg and Kohn~\cite{Hohenberg:1964p65}, the exact electronic ground-state energy can be formally obtained by the following minimization over one-electron densities $\rho$
\begin{align}
\label{eq:enerfunc.dft}
E = \min_\rho \left\{ F[\rho] + V[\rho] \right\},
\end{align}
where $F[\rho]$ is the universal density functional and $V[\rho] = \int \rho(\rr) v_{\mathrm{ext}}(\rr) d\rr$ is the energy associated with the external potential $v_{\mathrm{ext}}(\rr)$. The functional $F[\rho]$ can be expressed with Levy's constrained search over general wave functions $\Psi$ yielding the density $\rho$~\cite{Lev-PNAS-79}
\begin{equation}
\label{eq:universal.dft}
F[\rho] = \min_{\Psi \to \rho} \langle \Psi | \widehat T + \widehat V_{ee} | \Psi \rangle,
\end{equation}
where $\widehat T$ is the kinetic energy operator and $\widehat V_{ee}=\sum_{i<j} 1/r_{ij}$ the coulombic electron interaction operator.

In the range-separation scheme (see, e.g., Ref.~\onlinecite{Toulouse:2004p062505}), the electron interaction is decomposed as
\begin{align}
\label{eq:separation}
\widehat V_{ee} = \widehat V_{ee}^{\lr} + \widehat V_{ee}^{\sr},
\end{align}
where $V_{ee}^{\lr} = \sum_{i<j} {\mathrm{erf}(\mu r_{ij})}/{r_{ij}}$ is the long-range interaction, $V_{ee}^{\sr} = \sum_{i<j}{\mathrm{erfc}(\mu r_{ij})}/{r_{ij}}$ is the complement short-range interaction and $\mu$ is a parameter controlling the range of separation. The long-range interaction reduces to the Coulomb interaction at large interelectronic distances ($r_{ij} \gg 1/\mu$), while the short-range interaction reduces to the Coulomb interaction at small interelectronic distances ($r_{ij} \ll 1/\mu$).
Employing the error function makes the evaluation of the two-electron integrals simple, because there is an analytical formula for Gaussian basis sets. It also represents the most common choice in the literature. A long-range universal density functional $F^{\lr}[\rho]$ is then defined as
\begin{eqnarray}
F^{\lr}[\rho] = \min_{\Psi \to \rho} \langle \Psi | \widehat T + \widehat V_{ee}^{\lr} | \Psi \rangle,
\label{eq:lrf}
\end{eqnarray}
and the complement short-range density functional $F^{\sr}[\rho]$ is simply the remainder
\begin{equation}
\label{eq:srdft}
F^{\sr}[\rho] = F[\rho] - F^{\lr}[\rho].
\end{equation}

\begin{figure}
\includegraphics[scale=0.35]{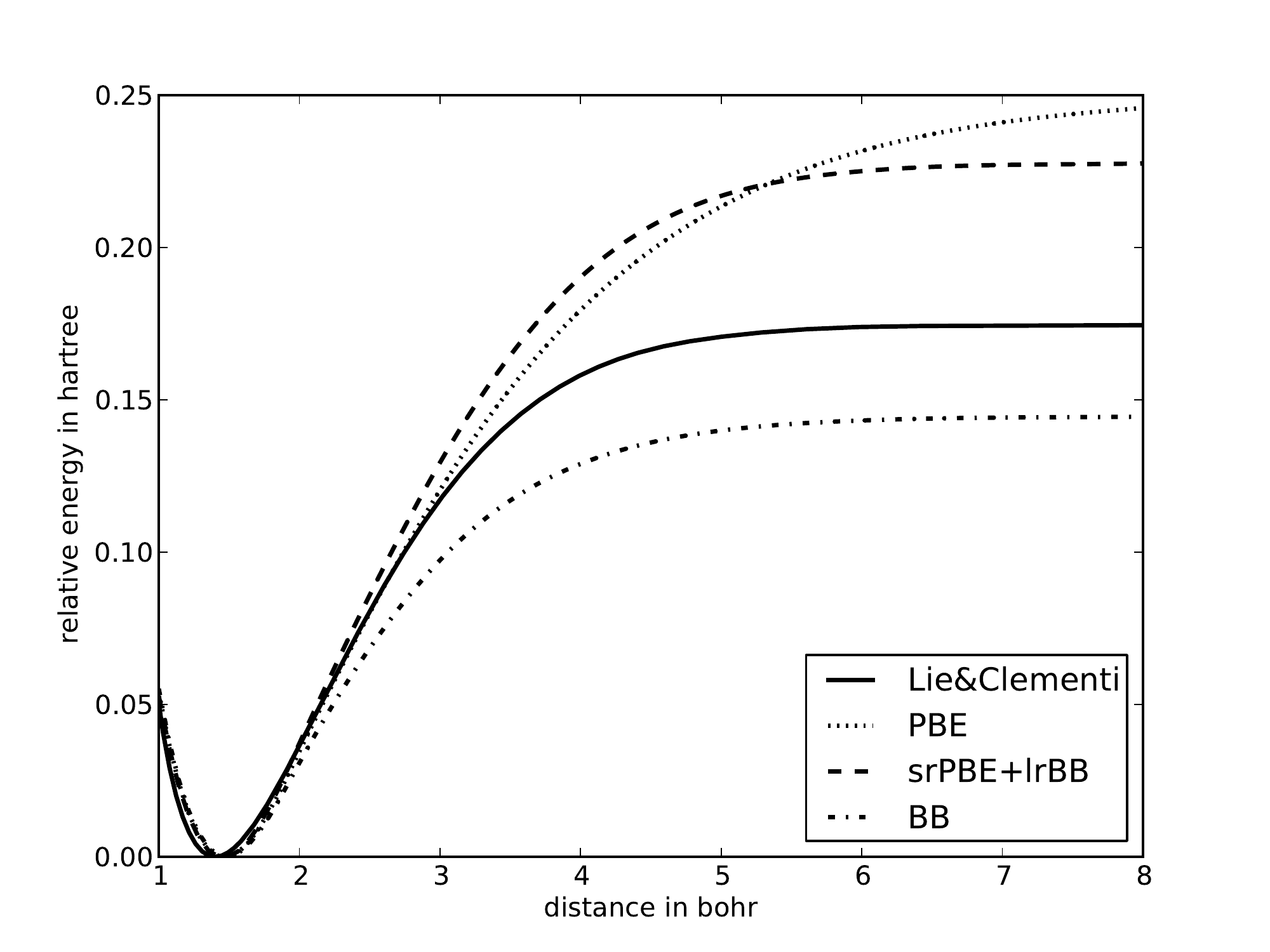}
\caption{\label{fig:h2}Dissociation curve of the H$_{2}$ molecule.}
\end{figure}

We reformulate Eq.~(\ref{eq:lrf}) as a constrained search over one-electron reduced density matrices $\Gamma$ yielding density $\rho$
\begin{align}
\label{eq:lrdft}
F^{\lr}[\rho] = \min_{\Gamma \to \rho} \{ T[\Gamma] + E_{ee}^{\lr}[\Gamma] \},
\end{align}
where $T[\Gamma]$ is the exact kinetic energy density-matrix functional
\begin{align}
\label{eq:kinetic}
T[\Gamma] =& -\frac{1}{2}\iint \delta(\rr - \rr') \, \nabla_\rr^2 \, \Gamma(\rr,\rr') d\rr d\rr',
\end{align}
and $E_{ee}^{\lr}[\Gamma]$ is the long-range electron interaction density-matrix functional defined with the following constrained search over wave functions $\Psi$ yielding $\Gamma$
\begin{align}
\label{eq:lree}
E_{ee}^{\lr}[\Gamma] =& \min_{\Psi \to \Gamma} \langle \Psi | \widehat V_{ee}^{\lr} | \Psi \rangle.
\end{align}
Combining Eqs.~\eqref{eq:enerfunc.dft},~\eqref{eq:srdft} and~\eqref{eq:lrdft}, we can re-express the exact ground-state energy as the following minimum over $N$-representable density matrices $\Gamma$
\begin{align}
E = \min_{\Gamma} \left\{ T[\Gamma] + V[\rho_{\Gamma}]+ E_{ee}^{\lr}[\Gamma] + F^{\sr}[\rho_{\Gamma}] \right\},
\end{align}
where $\rho_{\Gamma}$ is the density obtained from $\Gamma$. As usual, we can split up the long-range and short-range functionals into Hartree and exchange-correlation contributions, $E_{ee}^{\lr}[\Gamma] = E_{\H}^{\lr}[\rho_\Gamma]+E_{xc}^{\lr}[\Gamma]$ and $F^{\sr}[\rho_\Gamma]=E_{\H}^{\sr}[\rho_\Gamma]+E_{xc}^{\sr}[\rho_{\Gamma}]$, and after recomposing the total coulombic Hartree functional $E_{\H}^{\lr}[\rho_\Gamma]+E_{\H}^{\sr}[\rho_\Gamma]=E_{\H}[\rho_\Gamma]=(1/2) \iint (1/r_{12}) \rho_\Gamma(\rr_{1}) \rho_\Gamma(\rr_{2}) d\rr_{1}d\rr_{2}$, we finally obtain the srDFT+lrDMFT energy expression~\cite{Pernal:2010p052511}
\begin{eqnarray}
E = \min_{\Gamma} \left\{ T[\Gamma] + V[\rho_{\Gamma}]+ E_{\H}[\rho_\Gamma] + E_{xc}^{\lr}[\Gamma] + E_{xc}^{\sr}[\rho_{\Gamma}] \right\}.
\nonumber\\
\label{eq:enerfunc.dftlrdmft}
\end{eqnarray}
With the exact long-range density-matrix functional $E_{xc}^{\lr}[\Gamma]$ and the exact short-range density functional $E_{xc}^{\sr}[\rho]$, the minimum $E$ in Eq.~\eqref{eq:enerfunc.dftlrdmft} will be the exact ground-state energy. The minimizer $\Gamma$ will yield the exact ground-state density. However, it will not be the exact ground-state density matrix. 

In practice, approximations must be used for $E_{xc}^{\sr}[\rho]$ and $E_{xc}^{\lr}[\Gamma]$. The definition of $E_{xc}^{\sr}[\rho]$ is identical to that in the literature~\cite{Toulouse:2004p062505} and a number of approximations are available. We use the short-range PBE exchange-correlation functional of Ref.~\onlinecite{Goll:2005p2378}. The choice is based on a quick screen of short-range density functionals. Details will follow in a future publication. For $E_{xc}^{\lr}[\Gamma]$, we use the long-range BB functional~\cite{Pernal:2010p052511}, whose spin-summed expression in a real-valued orthonormal basis $\{\chi_a(\rr)\}$ is

\begin{figure}
\includegraphics[scale=0.35]{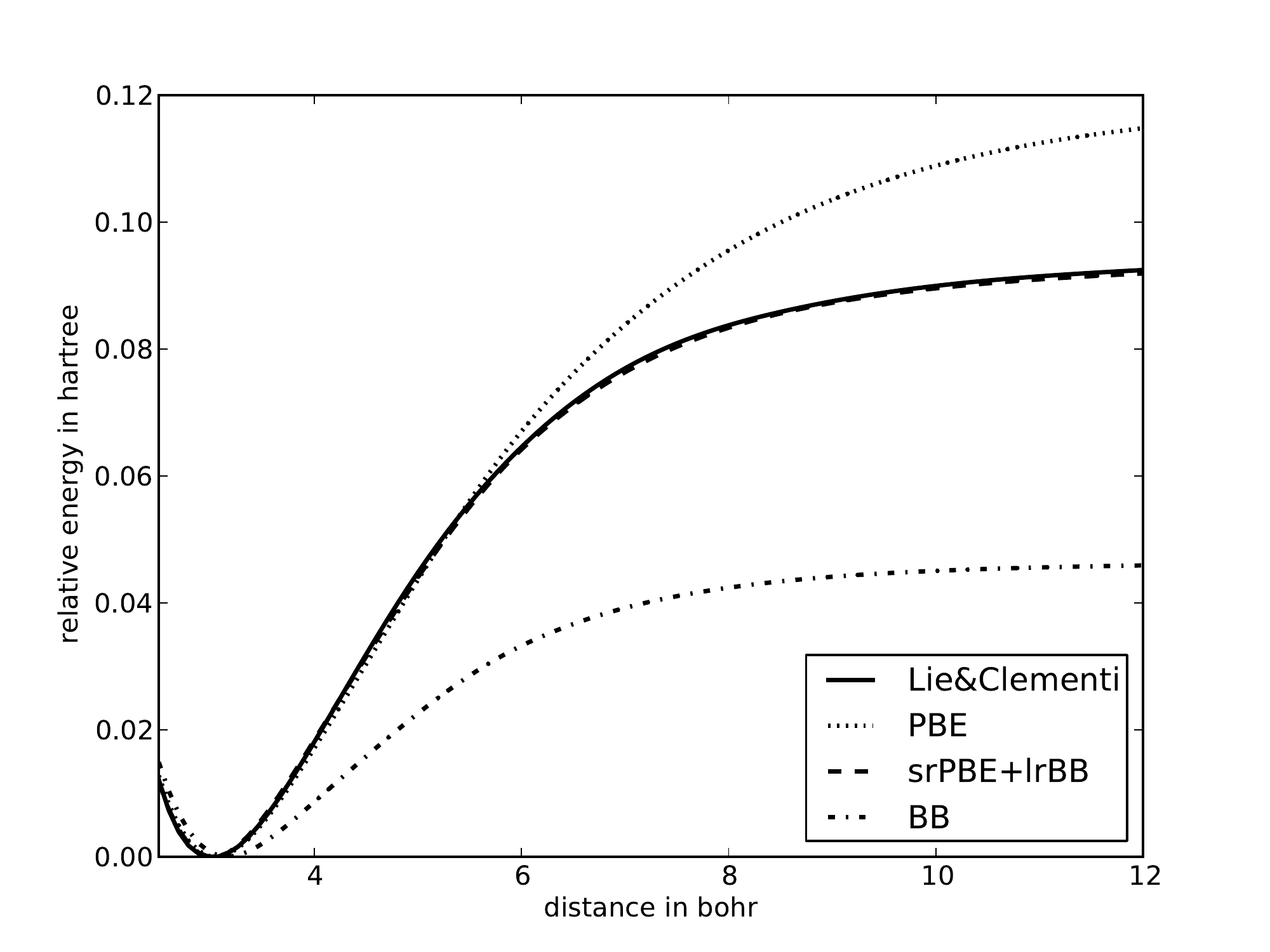}
\caption{\label{fig:lih}Dissociation curve of the LiH molecule.}
\end{figure}

\begin{align}
\label{eq:lrBB}
E_{xc}^{\text{lrBB}}[\Gamma] =& - \sum_{abcd} (\Gamma^{1/2})_{ab} (\Gamma^{1/2})_{cd} \langle ac|db \rangle^{\mathrm{lr}},
\end{align}
where $(\Gamma^{1/2})_{ab}$ are the elements of the square root of the matrix $\Gamma$ and $\langle ac|db \rangle^{\mathrm{lr}}$ are the two-electron integrals with long-range interaction $\mathrm{erf}(\mu r)/r$. The energy in the srPBE+lrBB approximation for closed-shell systems is thus calculated as
\begin{eqnarray}
\label{eq:srPBElrBB}
E &=& \min_{\Gamma} \Bigl\{ 2\sum_{ab} \Gamma_{ab} h_{ab} + 2 \sum_{abcd} \Gamma_{ab} \Gamma_{cd} \langle ac|bd \rangle 
\nonumber\\
&& - \sum_{abcd} (\Gamma^{1/2})_{ab} (\Gamma^{1/2})_{cd} \langle ac|db \rangle^{\mathrm{lr}} + E_{xc}^{\text{srPBE}}[\rho_{\Gamma}] \Bigl\},
\nonumber\\
\end{eqnarray}
where $h_{ab}$ are the one-electron integrals (kinetic + external potential), $\langle ac|bd \rangle$ are the two-electron integrals with full coulomb interaction needed for the Hartree energy, and $E_{xc}^{\text{srPBE}}[\rho_{\Gamma}]$ is evaluated with the density $\rho_{\Gamma}(\rr)=2\sum_{ab} \Gamma_{ab} \chi_a(\rr) \chi_b(\rr)$. The energy functional in eq.~\eqref{eq:srPBElrBB} is minimized over all \textit{N}-representable $\Gamma$. The \textit{N}-representability conditions read $\Gamma^{T} = \Gamma$, $\mathrm{Tr}(\Gamma) = N/2$ and $\Gamma^2 \leq \Gamma$, where \textit{N} is the number of electrons.

\vskip 2mm \noindent {\it Computational details.} \hskip 2mm
We have implemented the srPBE+lrBB method in our existing DMFT code, taking the one- and two-electron integrals from the Molpro quantum chemistry package~\cite{MOLPRO_brief}. The minimization is performed with the projected gradient algorithm~\cite{Cances:2008p2948}. It has been proven that the projected gradient algorithm is particularly efficient for the BB density-matrix functional. We choose a range-separation parameter of $\mu=0.4$ bohr$^{-1}$. This value was supported by some test calculations and it lies in the typical range of 0.3 to 0.5 used in the literature. A more detailed investigation will be published in a future paper. The standard BB calculations are performed with our DMFT code, and the standard Kohn-Sham PBE calculations are performed with Molpro. All calculations are done in a spin-restricted formalism. We used the cc-pVTZ basis for all systems. For the test of basis convergence on H$_2$, we also performed calculations with cc-pVXZ (X=D,T,Q,5)~\cite{Dunning:1989p1007}.

\begin{figure}
\includegraphics[scale=0.35]{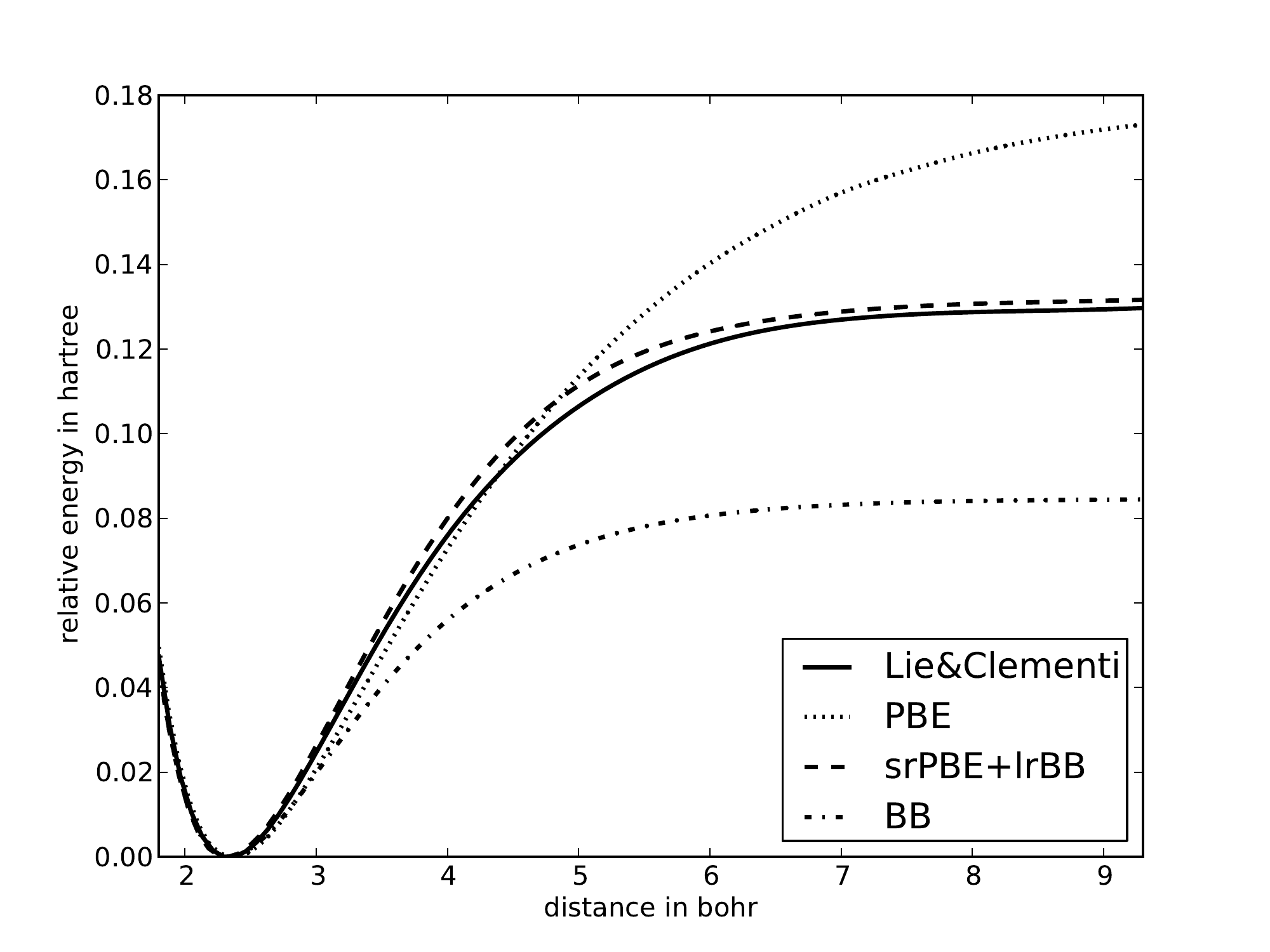}
\caption{\label{fig:bh}Dissociation curve of the BH molecule.}
\end{figure}

\vskip 2mm \noindent {\it Results.} \hskip 2mm
Figure~\ref{fig:bs} displays the convergence of the total energy of the H$_{2}$ molecule at equilibrium distance as a function of the basis sets cc-pVXZ (X=D,T,Q,5). The accurate total energy~\cite{Lie:1974p1275} is indicated with a solid horizontal line. We see the well known fact that density functionals are weakly basis-size dependent: the PBE energy (dotted curve) shows a fast basis convergence, the energy being already converged with cc-pVTZ up to 0.7 mhartree. In contrast, density-matrix functionals are much more basis-size dependent: the BB energy (dash-dotted curve) is not completely converged even with the cc-pVQZ basis, and going to cc-pV5Z basis the energy lowers by another 2 mhartree. Note also that the BB total energy is much too low. The srPBE+lrBB energy (dashed curve) displays a fast basis convergence. The curve runs virtually parallel to the PBE curve, and the energy is already converged up to 0.8 mhartree with the cc-pVTZ basis. The favorable basis dependence of srPBE+lrBB is not a surprise since short-range correlations, which determine the basis convergence, are efficiently described by a density functional in this approach.

First we discuss the prototype situation for static correlation, the dissociation curve of the H$_{2}$ molecule (see Fig.~\ref{fig:h2}). To facilitate comparisons, we show relative dissociation energy curves where the minimum for each method is set to zero. The accurate reference curve (solid curve) is from Lie and Clementi~\cite{Lie:1974p1275}. The PBE curve (dotted, shifted downward by -7.6 mhartree relative to the reference) is in excellent agreement with the reference around the equilibrium distance, where dynamic correlation dominates. However, for stretched bond distances, it shows a qualitatively incorrect behavior: the total energy does not yet saturate at a distance of 8 bohr. This is prototypical for the failure of common density-functional approximations in a spin-restricted Kohn-Sham formalism for describing static correlation. In contrast, the BB curve (dash-dotted curve, shifted upward by +17 mhartree relative to the reference) is qualitatively correct: the total energy saturates at a bond distance of about 5 bohr. However, the energy well is too shallow. The range-separated srPBE+lrBB curve (dashed curve, shifted by -4.9 mhartree relative to the reference) is virtually identical with the reference curve near the equilibrium like the PBE curve, and the total energy saturates at around 5 bohr like the BB curve. Nevertheless, the energy well remains too deep by 47 mhartree compared to the reference. It has been checked that this residual error is due to the short-range PBE approximation.

\begin{figure}
\includegraphics[scale=0.35]{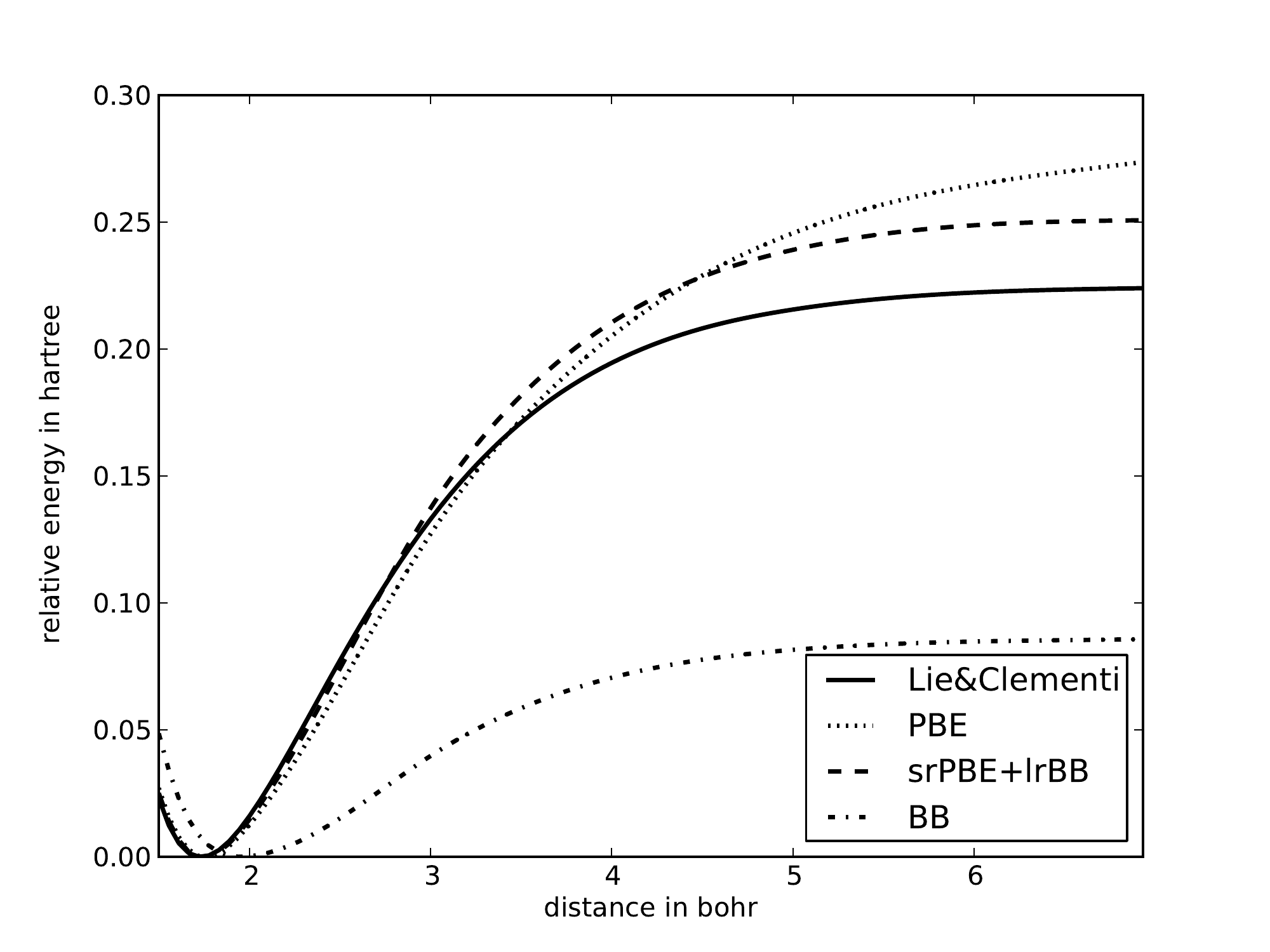}
\caption{\label{fig:hf}Dissociation curve of the HF molecule.}
\end{figure}

Figures~\ref{fig:lih},~\ref{fig:bh} and~\ref{fig:hf} show the relative dissociation energy curves for the LiH, BH and HF molecules. The solid curves represent accurate reference energies~\cite{Lie:1974p1275}. For LiH, the PBE, srPBE+lrBB and BB curves were shifted relative to the reference curve by -24, -28 and +39  mhartree, respectively. For BH, they were shifted by -45, -50 and +182 mhartree, respectively, and for HF by -64, -51 and +75 mhartree, respectively. A similar picture is found for all three molecules. Standard PBE performs well around the equilibrium distance, but the energy unphysically increases at large distances. The BB energy correctly saturates at a similar distance as the reference curve, but the curvature at equilibrium is underestimated and the energy well is too shallow. The range-separated srPBE+lrBB method inherits the good performance of PBE near the equilibrium and the correct behavior of BB at large distance. It is in very good agreement with the reference at all bond distances.

\vskip 2mm \noindent {\it Conclusions.} \hskip 2mm
The presented srPBE+lrBB method is a promising approach for accurately describing both dynamic and static electron correlation, as shown for bond dissociation curves of some test molecules. Around the equilibrium distance, it is as accurate as standard DFT with the PBE approximation, indicating a correct description of dynamic correlation. At stretched-bond distances, the total energy correctly saturates like DMFT with the BB approximation, indicating an adequate description of static correlation. This is achieved without artificial breaking of spin symmetry, and at a computational cost comparable to that of standard density functional approximations. In the future, we will explore how srPBE+lrBB performs for larger systems.

\vskip 0.5mm \noindent {\it Acknowledgements.} \hskip 2mm
D.R. acknowledges financial support from the German Research
Foundation (DFG) under grant number RO 3894/1-1. K.P. acknowledges 
support from the Polish Ministry of Science and Higher Education (Grant No. N N204 159036).
Useful discussions with Andreas Savin are gratefully acknowledged. 

\bibliography{Library}

\end{document}